# Electrophysiological indicators of gesture perception

Maria E. Cabrera[1], Keisha Novak[2], Dan Foti[2], Richard Voyles[3], Juan P. Wachs[4*]

[1] School of Computer Science and Engineering, University of Washington, Seattle, WA, USA

[2] Department of Psychological Sciences, Purdue University, West Lafayette, IN, USA

[3] School of Engineering Technology, Purdue University, West Lafayette, IN, USA

[4] School of Industrial Engineering, Purdue University, West Lafayette, IN, USA

*Corresponding Author:

Juan P. Wachs, Ph.D.

Associate Professor,

James A. and Sharon M. Tompkins Rising Star Professorship

Regenstrief Center for Healthcare Engineering

Adjunct Professor of Surgery, IU School of Medicine

School of Industrial Engineering

Purdue University

315 N. Grant Street

West Lafayette, IN 47907

2
**Abstract**

**Background:** While there has been abundant research concerning neurological responses to gesture generation, the time course of gesture processing is not well understood. Specifically, it is not clear if or how particular characteristics within the kinematic execution of gestures capture attention and aid in the classification of gestures with communicative intent. If indeed key features of gestures with perceptual saliency exist, such features could help form the basis of a compact representation of the gestures in memory.

**Methods:** This study used a set of available gesture videos as stimuli. The timing for salient features of performed gestures was determined by isolating inflection points in the hands' motion trajectories. Participants passively viewed the gesture videos while continuous EEG data was collected. We focused on mu oscillations (10 Hz) and used linear regression to test for associations between the timing of mu oscillations and inflection points in motion trajectories.

**Results:** Peaks in the EEG signals at central and occipital electrodes were used to isolate the salient events within each gesture. EEG power oscillations were detected 343 and 400ms on average after inflection points at occipital and central electrodes, respectively. A regression model showed that inflection points in the motion trajectories strongly predicted subsequent mu oscillations ($R^2 = 0.961, p<.01$).

**Conclusion:** The results suggest that coordinated activity in the visual and motor cortices are highly correlated with key motion components within gesture trajectories. These points may be associated with neural signatures used to encode gestures in memory for later identification and even recognition.

# Acknowledgments

This work was supported by the Office of the Assistant Secretary of Defense for Health Affairs under Award No. W81XWH-14-1-0042 (JPW is Principal Investigator). Opinions, interpretations, conclusions and recommendations are those of the author and are not necessarily endorsed by the Department of Defense.

# 1. Introduction

Performing gestures is a form of communication that is used to complement, emphasize, or accompany verbal messages, or alternatively serve as an integral modality of communication on its own – without the verbal component. The meaning of a gesture may differ across cultures and societies, yet the perception and execution of gestures are intrinsic parts of human behavior; as such, there are cognitive processes associated with gesture perception, production and recognition. The relationship between perception and production of an action comes from the human mirroring system (Decety and Grèzes, 1999), activating regions of the brain involved in performing an action even when said action is only observed (Gazzola and Keysers, 2009). These activations are related with the potential to facilitate the interpretation and understanding of actions performed by others (Rizzolatti and Sinigaglia, 2010). This fundamental recognition is essential for survival, as well as social interactions and functioning. Specifically, empathy, communication, social skills, and overall coordination can all be traced back to the importance of motor representations (Urgen et al., 2013).

The purpose of this work is to analyze neural signatures from electroencephalographic (EEG) data recorded while observing gestures, to determine the relationship between key motion components of each gesture with oscillations in the mu frequency band, associated to the motor cortex. The present work is an extension of the analysis presented in Cabrera et al. (2017), in which a preliminary analysis was conducted on a reduced dataset to determine the existence of a relationship between salient kinematic points in observed gestures with oscillations in EEG potentials associated to both motor and visual cortices activation. The current work focuses on extending the analysis in order to corroborate this relationship. This is done by further increasing the sample size and also by expanding the number and type of analyses conducted.

Gesture processing and the human Mirror Neuron System (hMNS) activation have been examined previously in electrophysiological studies using mu suppression. Decreases in power within the mu frequency band (8-13 Hz) measured at central electrodes (Braadbaart et al., 2013), and considered maximal at central electrodes directly above the motor cortex (Muthukumaraswamy et al., 2004), suggest motor cortex activation not only during action execution but also observation. Mu suppression is thought to reflect desynchronization in motor cortical activity and has been argued to play a role in contexts involving social interaction (Perry et al., 2011) as well as passive action observation (Hogeveen et al., 2014). In the same frequency band, EEG activity at occipital electrodes has been associated with activation of the visual cortex and visual processing (Neuper et al., 2005; Pfurtscheller et al., 1994).

Engel et.al (2008) attempted to answer the question whether it is simply the movement trajectory (a visual feature) or the task goal (the intention of the observer) that is crucial for hMNS activation using functional magnetic resonance imaging (fMRI). Their findings suggest that it is not a specific visual feature that activates the hMNS, rather the hMNS seems to respond when an observed movement is matched to a motor representation triggered by the intentional goal of the observer (Engel et al., 2008). This aspect is clearly distinct from purely processing visual features. Such a correlation to movement execution can be constructed from both biological and non-biological (i.e. robotic arms or virtual avatars) stimuli and it is conceivable that this may often arise implicitly or subconsciously, when visual features trigger strong associations with human movements.

In addition to Engel's work, there is abundant literature testing the described effect of gestures on neural activity. Quandt et al. (2012) showed that gesture performance had an effect on alpha oscillations through recorded EEG. Although alpha oscillations within the same frequency band as mu oscillations, mu refers specifically to activity over the motor cortex (i.e., central scalp electrodes), whereas alpha is not specific to any one region. Their study included different gesture types, such as deictic (i.e. pointing gestures) and iconic (i.e. pantomime, related to speech). Their results suggest that different types of gestures engage the motor system in different ways based on EEG measurements on the alpha band. Quandt et al. (2012) associated the higher alpha oscillations on iconic gestures to the closeness of the meaning of the gesture to the action and the richer information. Wu and Coulson (2005) used EEG recordings and event-related potentials to determine the meaning of iconic gestures. These findings suggest that iconic gestures are subject to processes analogous to those evoked by other meaningful representations, such as pictures and words.

Dahan and Reiner (2016) studied gesture motions performed by humans vs non-humans (artificial objects), and also meaningful vs meaningless gestures. They found that the hMNS only activates with motions done by the kinematics of the human body. However, it is not clear which part of the stimulus is the cause for triggering the activation in the hMNS: the limb-like Gestalt of the moving device, the movement trajectories that resemble biological trajectories, or, as suggested by Gazzola et al. (2007), the understanding of the action goal of an observed action, even when a robot arm's movements are used as stimuli. One reason could be that the trajectories of the objects were processed as biological movements, because they followed a smooth path.

A previous study compared central mu activity and occipital alpha activity observed in different gesture types (Streltsova et al., 2010). They reported central mu activity but not occipital alpha activity when the gestures observed were meaningless and not communicative. Streltsova's study considered the averaged EEG power over 1000ms bins. Fluctuations in mu activity while observing communicative gestures is not necessarily related to the total amount of movement;

rather, it is thought to correlate to the processing of gesture meaning. Building on this finding, we tested whether the timing of mu oscillations during gesture observation would follow a unique time course for each gesture. However, we measured mu power continuously, allowing us to extract the time of mu oscillations in a more fine-grained manner; those peaks in mu activity could be used to infer the time that most salient gesture characteristics occur. We hypothesize that these characteristics would correspond to inflection points in the kinematic trace of the gesture manifestation. The potential impact of this finding is that we could use these inflection points to represent the most salient aspects of gestures, so gestures with the same inflection points will be perceived as similar and those with different inflection points will be seen as different gestures.

2. Results

   2.1. Descriptive Analysis

   The acquired EEG responses were averaged across participants for each gesture class to test if a pattern can be extracted from commonalities in mu activity. The Occipital Cluster showed a decrease in EEG power at 10 Hz (

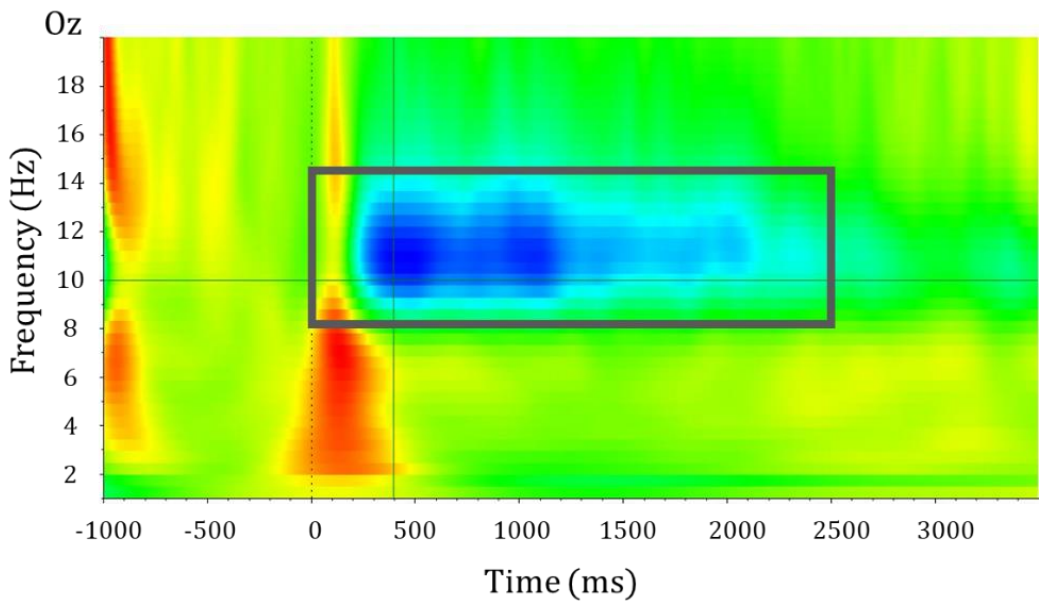

Figure 1). This is consistent with corresponding alpha activity and signals visual cortex activation associated with viewing a stimulus. This response was sustained, averaging over all gesture classes, within the 400-2500ms time range, indicating continuous visual activity for the period of time in which participants attended to the gesture videos and highlighted with a gray box

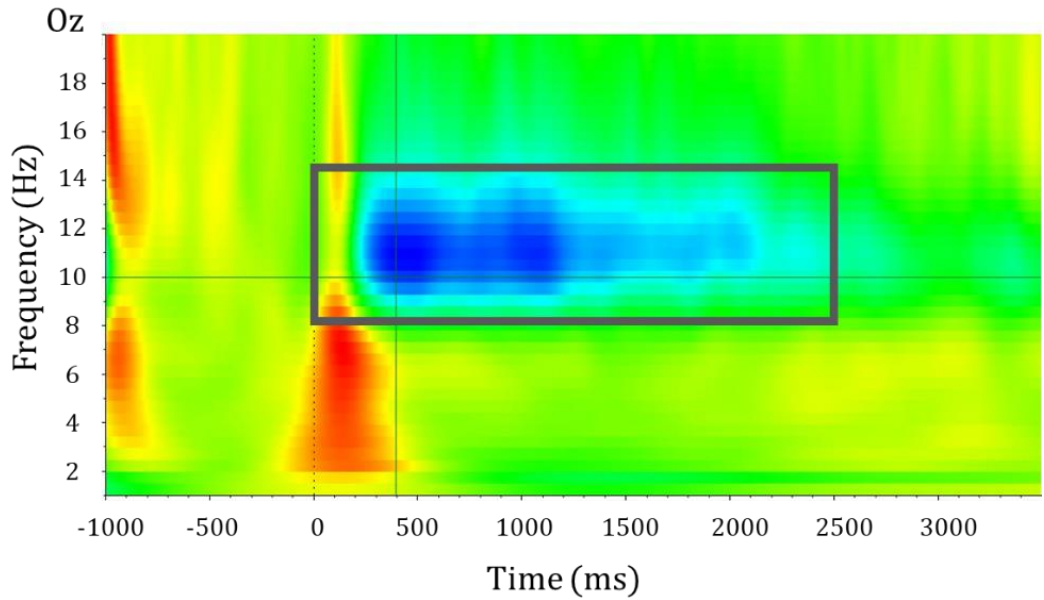

in

Figure 1.

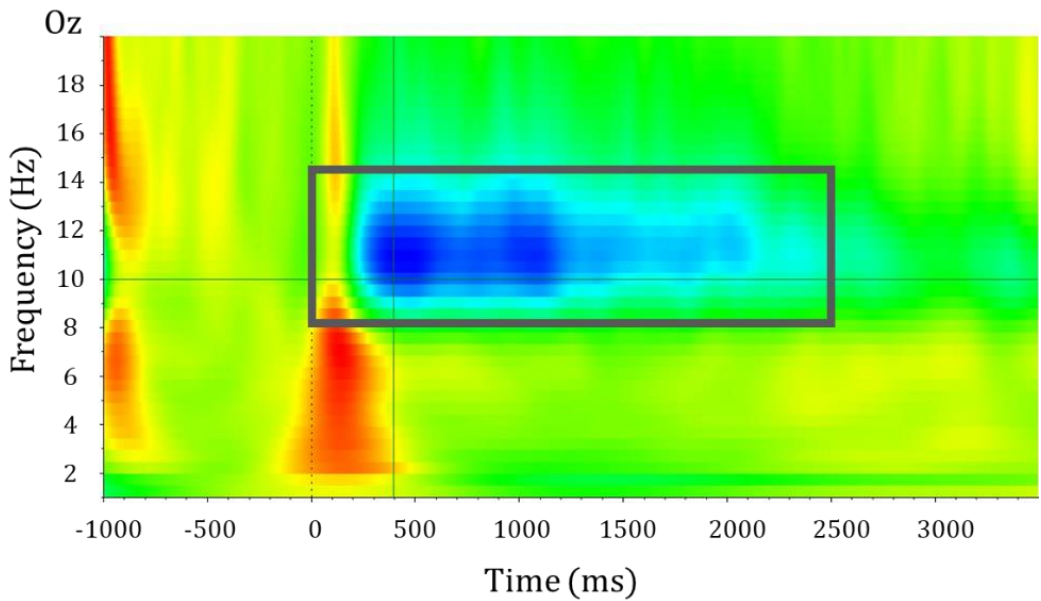

*Figure 1*. EEG power at occipital electrodes (Oz), averaged across all gesture stimuli (x-axis=time, y-axis=frequency; blue=decrease in power, red=increase). The gray box indicates a sustained decrease in alpha power corresponding to visual processing.

Regarding the Central Cluster, a different pattern was observed. There was evidence of mu suppression along the 10 Hz band between 200ms and 2500ms, within the gestures' duration. The most accentuated cluster of mu suppression occurred between 200-600ms with a peak at 400ms,

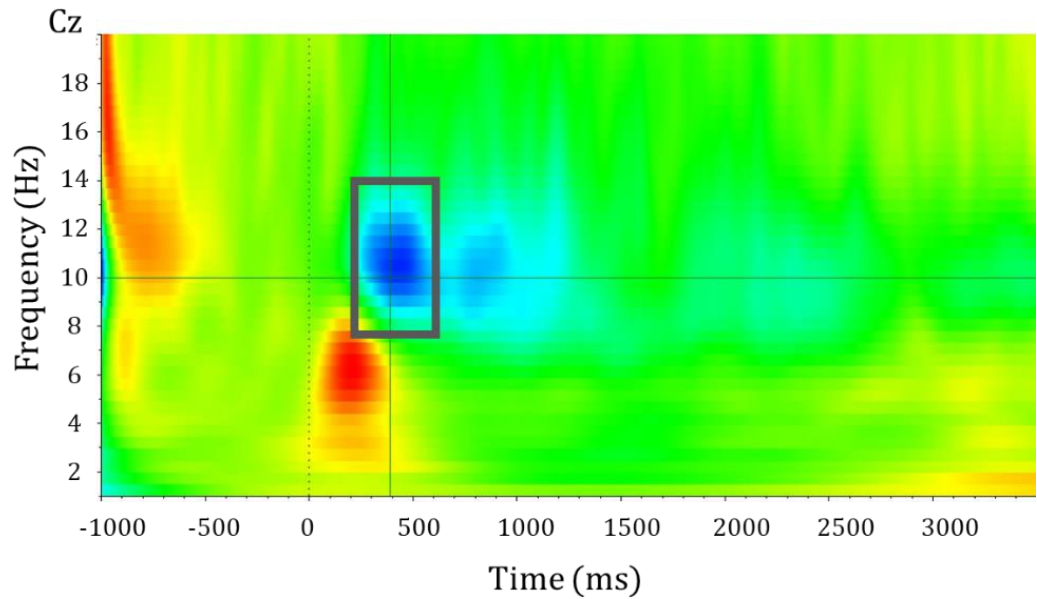

highlighted in

Figure 2 by a gray box. This indicates an automatic suppression of mu activity, independent of any salient characteristic of each gesture class. Expanding this observation, 15 out of the 20 gesture classes (75%) showed a mu peak even before any inflection points were registered in the gesture class with an average response of 300ms. This is consistent with previous results found in a preliminary study reported by Cabrera et al. (2017). Additionally, a previous EEG study also shows mu suppression within the first second of gesture viewing, regardless whether the gesture was communicative or meaningless in nature; this is a general response within the motor cortex distinct from visual cortical response apparent at occipital electrodes (Streltsova et al., 2010).

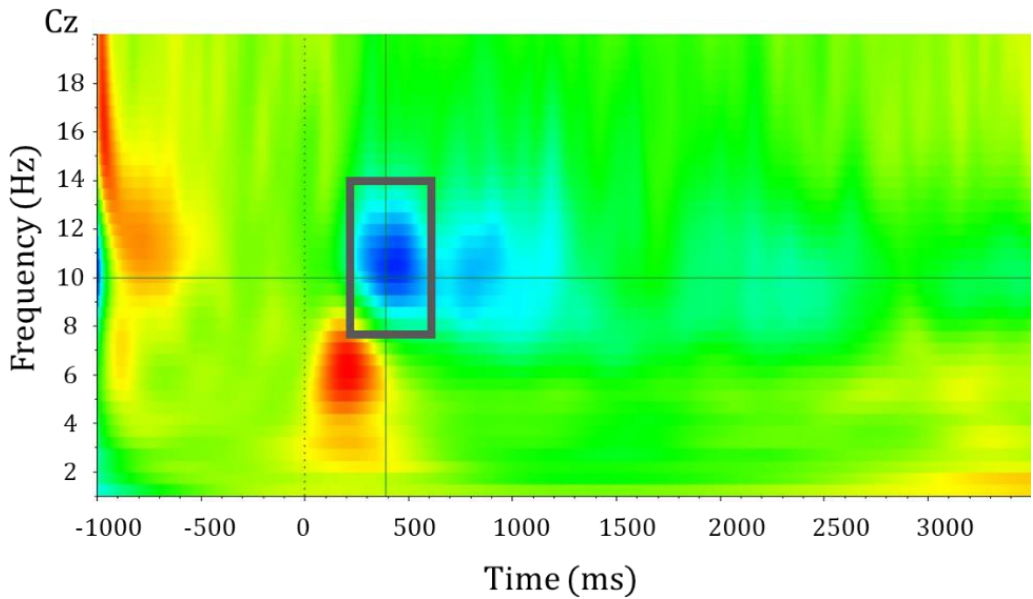

*Figure 2.* EEG power at central electrodes (Cz), averaged across all gesture stimuli (x-axis=time, y-axis=frequency; blue=decrease in power, red=increase). The gray box indicates mu suppression corresponding with gesture processing.

Breaking down the data per each gesture class, 85 peaks for mu activity were detected, defined as local maxima occurring after stimulus onset (counted after the first IP occurred in each gesture class). These positive-going peaks of mu activity represent the first evidence of dynamic mu oscillatory activity associated with gesture interpretation. All gesture classes elicited a first peak in mu activity with an associated average latency of 308ms (SD = 179ms) between IP and mu peak occurrence.

For all 66 IP occurrences considered in the stimulus set, there was a mu peak correspondence with an average time interval of 352ms (SD = 94ms). There were 19 peaks in mu activity that did not correspond directly with an IP occurrence: 12 of them occurred after the last IP had occurred, while 7 (8.2% of all mu peaks) were elicited in between IP, within the time interval the gesture was being executed (these were found in only 6 of the gesture classes). The average latency for these not-corresponded peaks was 972ms. Given the nature of the gesture vocabulary, each gesture began and ended with the performer doing the same resting pose: arms extended with the hands at hip level. The occurrence of a late mu peak (60% of the gesture classes) could be attributed to recognizing the resting pose at the end of the executed gesture.

    Next, we extended the descriptive analysis of the results to gesture class sub-groups according to number of inflection points found within. Regarding gesture classes with only 2 IP, 2 out of 3 (66.7%) presented two mu peaks after the first IP and before the second, and all classes presented an elicited mu peak towards the end of the gesture execution. The first elicited mu peak could correspond to the first IP with an average occurrence latency of 396ms (SD = 151ms), while the interval between the IP and the second mu peak averaged 925ms. From the gesture sub-groups with 3 IP and 4 IP, only two gesture classes (29%) elicited non-corresponding mu peaks in between IP occurrences respectively. The average latency observed between IP and mu peak occurrence was 339ms in average (SD = 88ms) for the group with 3 IP and 336ms (SD = 40ms) for the group with 4 IP. The only gesture class with 5 IP elicited the same number of mu peaks as IP with an average latency of 448ms (SD = 93ms). A pattern seems to appear: when there are not many salient features coming from motion (number of observed IP), additional mu elicitation is found more frequently in such gestures.

Two gesture examples are examined in more detail, namely G7 and G12.

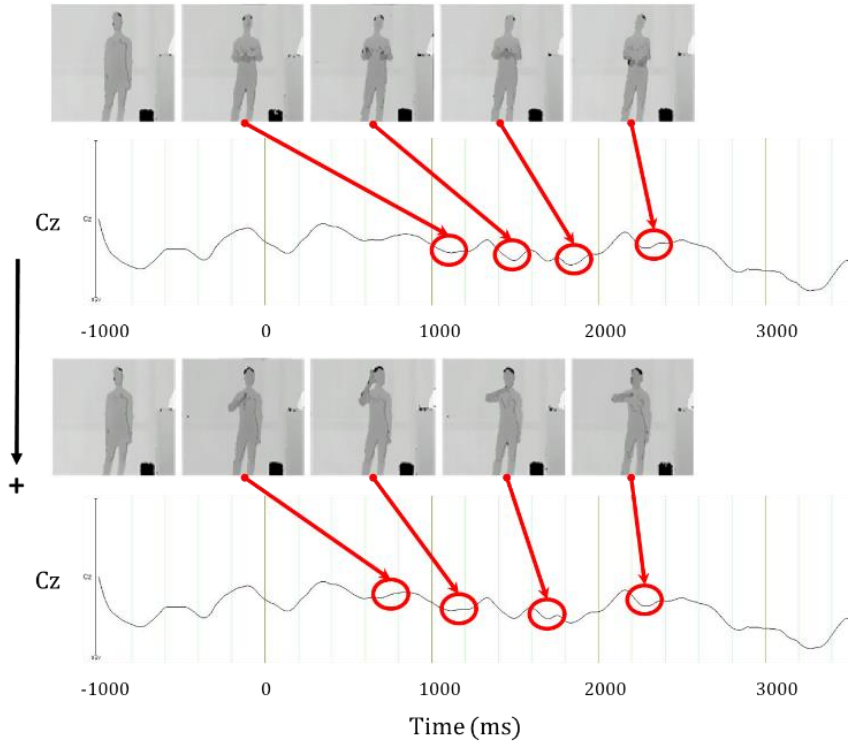

Figure 3 shows the time course of mu activity for each gesture, using the 10 Hz wavelet. Frames from the gesture video corresponding with the IP found showed above each graph are mapped to the signal time series with indicative arrows. The corresponding mu peaks are labelled with a red circle. Note that positive power is plotted down in both mu graphs. The latencies were 426ms for G7 and 348ms for G12, respectively.

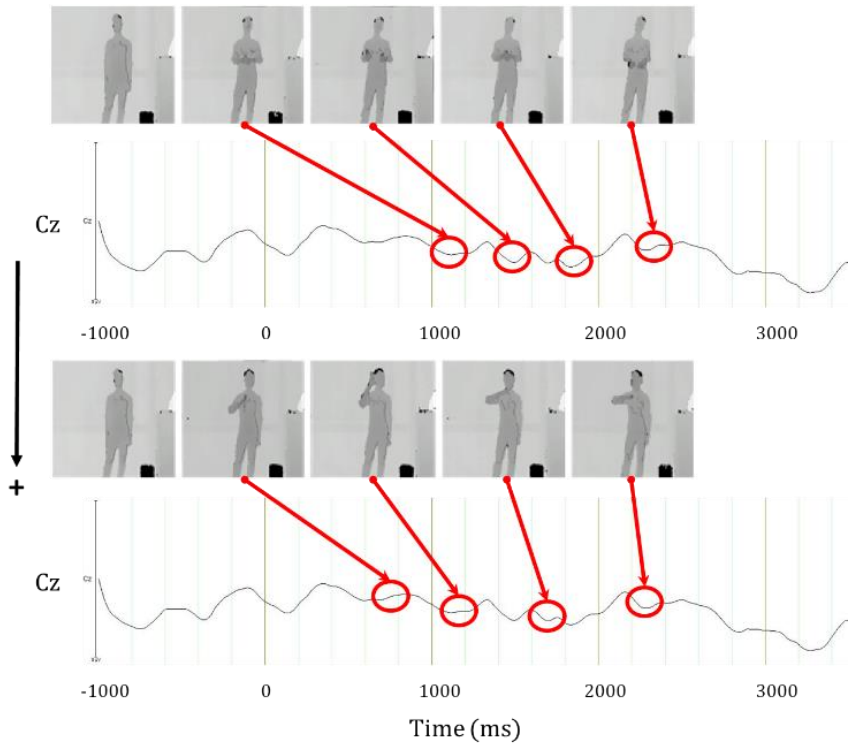

*Figure 3.* Time course of mu activity (10 Hz wavelet layer) for two representative gestures, each of which had four inflection points, shown in the embedded video frames.

These two gestures were used to compare against previous preliminary results reported (Cabrera et al., 2017). These two gestures also showed correspondence previously between the number of inflection points and the observed peaks in mu activity. These results continue to be consistent with the proposition that inflection points function as placeholders within gesture expression, and that these placeholders modulate activity within the motor cortex that is associated with gesture processing and interpretation.

## 2.2. Statistical Analysis

### 2.2.1. Regular Analysis

For this analysis, the time occurrence of an inflection point in motion trajectory were used for the abscissa axis while the time occurrence of a peak in EEG information, either in the averaged Central Cluster corresponding to activation in the motor cortex (

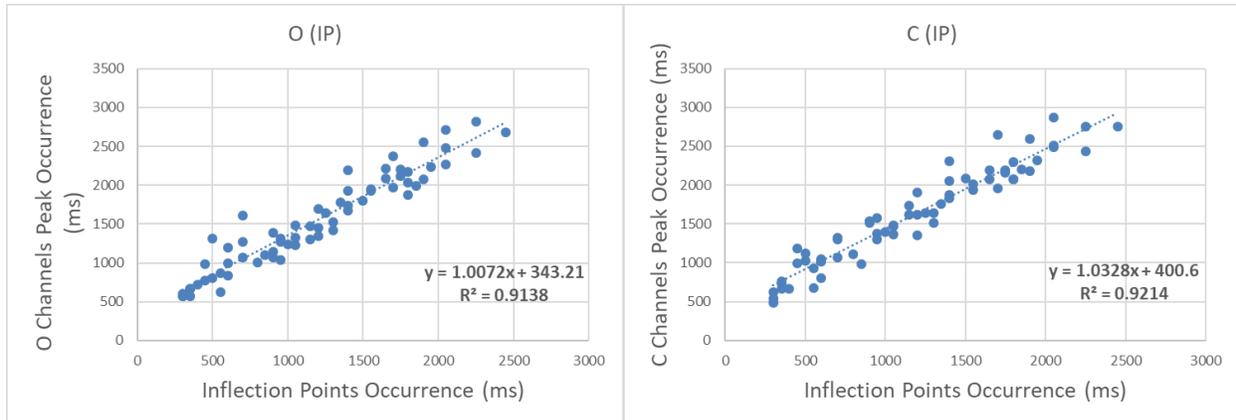

Figure 4, a) or the averaged Occipital Cluster corresponding to activation in visual cortex

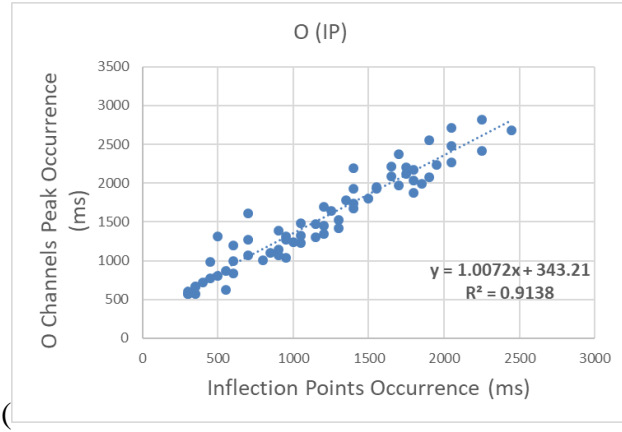

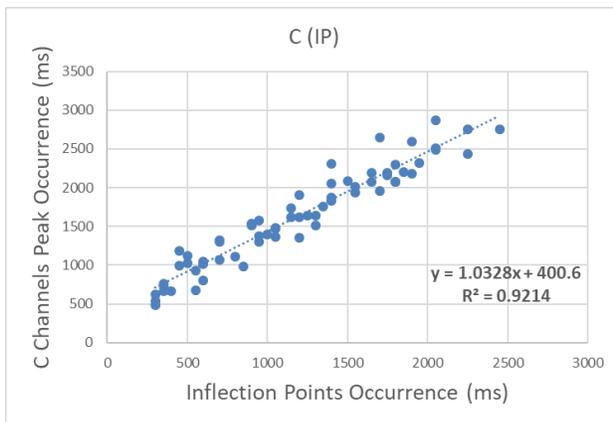

Figure 4, b). Lines were fitted and the coefficients of determination $R^2$ were obtained for this scenario, $R^2 = 0.9214$ and $R^2 = 0.9138$ respectively ($p < 0.001$) per channel type.

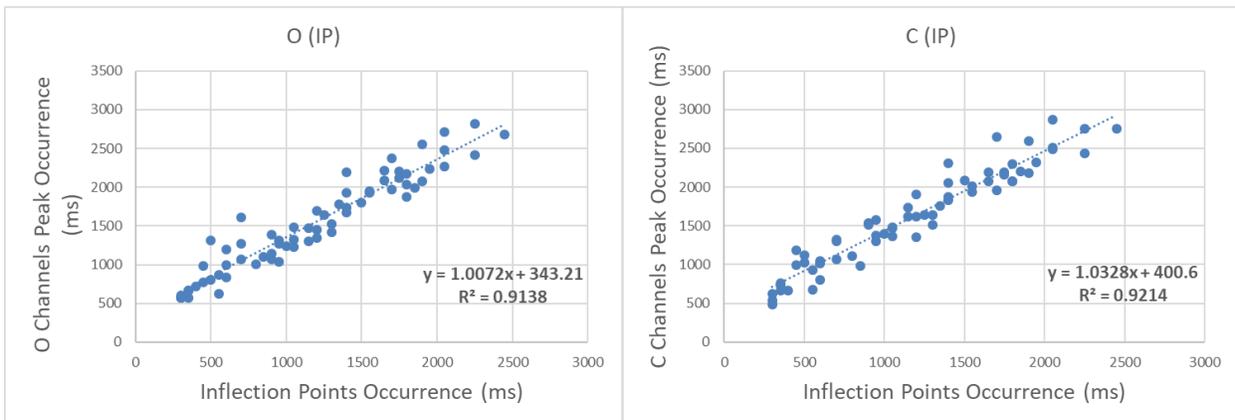

*Figure 4.* Regular analysis: EEG peak occurrence as a function of inflection points in motion at the Central Cluster (mu activity; left) and at the Occipital Cluster (alpha activity; right).

A high correlation was found between the incidences of salient motion components with peaks in neural oscillations. Another result worth highlighting from these fitted lines are the independent terms obtained, associated with time lags or reaction time between the motion stimuli and the EEG response. In the case of the Occipital Cluster the lag was found at $t_O = 343.21\ ms$, while the lag for the Central Cluster was found at $t_C = 400.6\ ms$. An additional analysis was conducted, acknowledging the fact that several IP with EEG spikes are nested within the same gesture. By considering only the occurrence of the 1st inflection point and the 1st EEG peak for the Central Cluster, a similar behavior is observed with the regression model showing a lag of $t_C^* = 377.13\ ms$ with $R^2 = 0.7636$.

### 2.2.2. Reverse Analysis

For this analysis the time occurrences of inflection points in gesture trajectories were plotted as a function of EEG peaks occurring in the averaged Central Cluster, as shown in

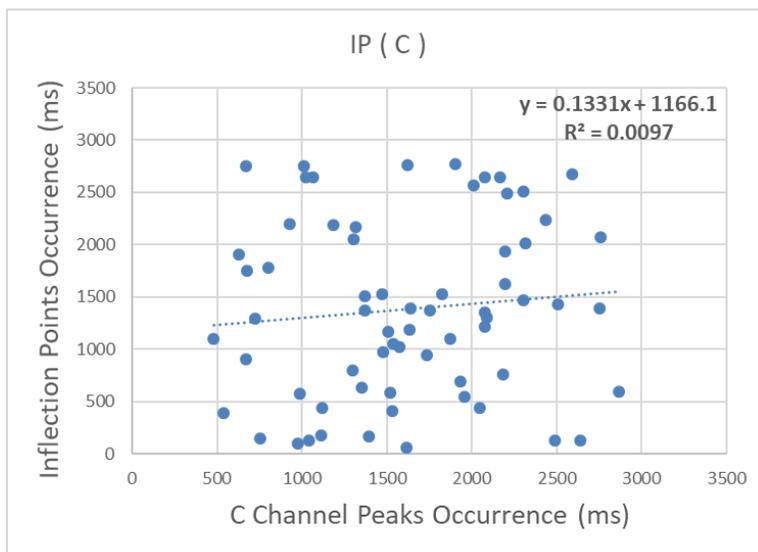

Figure 5. The fitted trend line and $R^2$ coefficient obtained from this analysis show no correlation ($R^2 = 0.0097$). No statistical significance was found for the Central Cluster as predictors for inflection points in motion in this instance ($p = 0.1353$). This analysis contests the idea of correlation simply due to relating contiguous measurements of IP against EEG, given the false reciprocal. The absence of correlation provides a basis to reject the assumption that EEG oscillations precede the occurrences of inflection points in gesture trajectories.

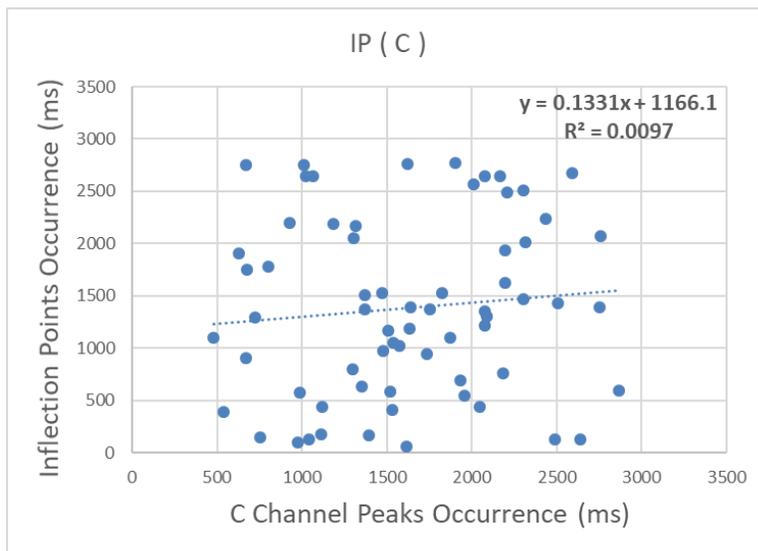

*Figure 5.* Reverse analysis: Occurrence of inflection points in motion as a function of EEG Peak occurrence (Central Cluster).

### 2.2.3. Random Analysis

In this analysis, the plotting scheme was the same as the regular analysis, showing Central Cluster peak occurrences as a function of inflection points in motion in

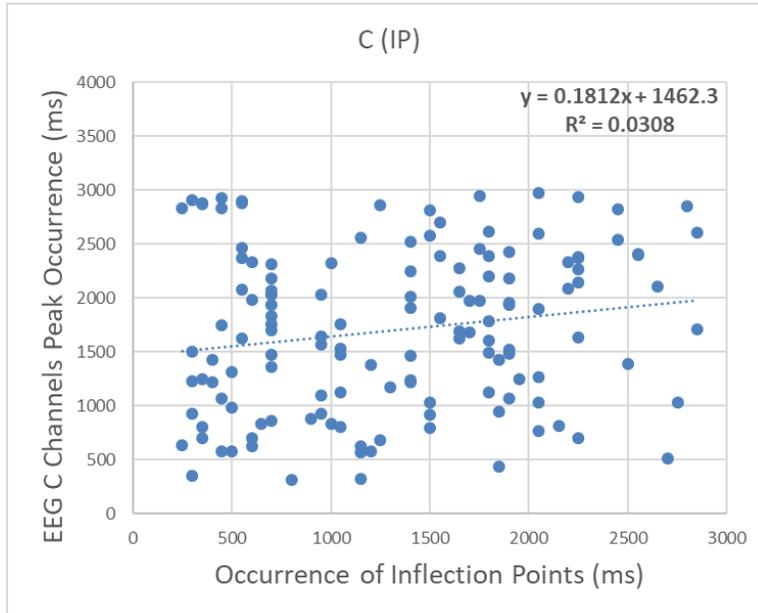

Figure 6. The fitted trend line and $R^2$ coefficient obtained from this analysis show no correlation ($R^2 = 0.0308$, $p = 0.6105$). The random nature in the EEG data used for this analysis showing no correlation with salient points of the gesture trajectories, provides strength to the previous results found with the regular analysis discarding a spurious patterning.

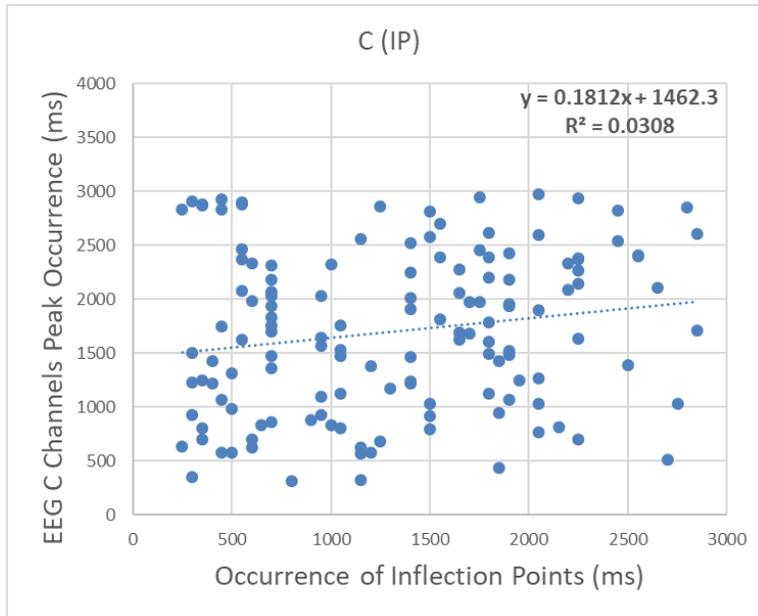

*Figure 6.* Random Analysis: EEG Central Cluster (flipped in time and shuffled by gesture) peaks occurrences as a function of inflection points in motion

### 2.2.4. Multi-level Hierarchical Analysis

The conducted analysis evaluated the change in $R^2$ for models predicting the occurrences of peaks in EEG Central Cluster ($EEG_C$) using the occurrences of inflection points used in the regular analysis ($Motion_{reg}$) and the reverse analysis ($Motion_{rev}$). Pearson correlation was found at 0.96 for $Motion_{reg}$ and -0.051 for $Motion_{rev}$. Additional results related to the fitted models and their significance are shown in Table 1.

*Table 1. Hierarchical multi-level regression results.*

| Variable | B | SE B | β | ΔR² |
|---|---|---|---|---|
| Step 1 |  |  |  | 0.00 |
| Motion_rev | -0.04 | 0.11 | -0.05 |  |
| Step 2 |  |  |  | 0.92 |

| | | | |
|---|---|---|---|
| Motion$_{rev}$ | -0.03 | 0.03 | -0.04 |
| Motion$_{reg}$ | 1.03 | 0.04 | **0.96*** |

The results show that the model predicting the EEG Central Cluster information becomes significant only when the inflection points used in the regular analysis ($Motion_{reg}$) are included as predictors.

### 3. Discussion

The results presented in this work are consistent with previous results reported in literature, showing that mu rhythms are sensitive to gesture observation (Braadbaart et al., 2013; Quandt et al., 2012). While previous studies have focused on the average reduction of EEG power, particularly mu suppression, while the stimulus is presented, this work focused on the oscillatory nature of mu activity and determining whether the occurrence of peaks in mu power would correspond to salient characteristics found in the gesture execution. The mentioned gesture characteristics are the points in the motion trajectory that display abrupt changes in either magnitude or orientation, namely the inflection points. These abrupt changes are considered relevant, connecting consecutive gesture phases (Kita et al., 1998). A linear relationship was found between the times when the gesture executed exhibited an inflection point and the time occurrences of peaks in EEG signals, particularly from the central and occipital electrodes. Average lags of 343ms for the occipital electrodes and 400ms for the central electrodes were found. These results show a close correspondence between gesture motion and EEG oscillations across the time of respective gesture execution and perception.

The detected oscillatory patterns mention both central and occipital electrodes. One main difference between the Streltsova study (2010) and ours was they considered average EEG power in 1000ms intervals, while we used precision wavelets to detect peaks within the EEG waveform. Close responses were found between the timing of peaks at central and occipital electrodes with the timing of inflection points in gesture motion trajectories. This could be interpreted as coordinated activation of visual and motor cortices involved in gesture processing. Given the earlier timing of the peaks found in occipital electrodes (343ms) to the ones found in central electrodes (400ms), this suggests visual cortical activity was driving motor cortical activity, although this needs to be verified in future research.

A descriptive analysis of the results found further describes the close correspondences between inflection points in motions with peaks in EEG signals associated to motor and visual cortices, focusing more on the differences among gesture classes. All inflection points extracted from gesture motion were matched to Central Cluster oscillations. There were additional oscillations in the Central Cluster that did not correspond to a preceding inflection point and occurred before the next in 6 of the gesture classes. Only 8.2% of all mu peaks found averaging over all 17 participants, displayed such behavior with an average lag of 972ms. These oscillations in the Central Cluster may be attributed to other salient characteristics of the gestures that were not considered in this work: hand shape configuration or relative position to the gesturer's body, among other features relevant in sign language. In regards to the peaks found during the gesture, in between IP, there could be other salient characteristics in those 6 gesture classes that were not related to the gestures' trajectories, for instance an additional response related to attention or learning which also manifest in the same band of 10 Hz oscillations (Başar, 2010). Additionally, 12 gesture classes exhibited a mu peak towards the end of the gesture video, when the gesturer was returning to a resting pose after executing each gesture class, which did not correspond to any extracted inflection points in motion. One interpretation of this could be related to recognizing the resting pose itself.

The conducted analysis paves the way towards future on gesture recognition. The EEG oscillations found related to salient points in motion can be used to build a global model of a compact representation of a gesture. This "mental snapshot" of the gesture includes rich information about the gestures' modes of variation. By "unfolding" this representation, human-like perception and movement production could be transferred to machines, making human-machine interaction more intuitive and natural. Understanding how people learn to mimic and produce gestures and transferring those techniques to machines can be considered a multidisciplinary effort, requiring expertise from psychology, motor control, human-machine interaction, machine learning and linguistics. The current findings are limited, among other things, by the salient points considered to describe the gesture's model; the use of motion trajectories can be used for gestures involving motion in the upper limbs, however the configuration of the hand or changes in the position of the fingers are overlooked presently. There is another limitation related to the communicative nature of the gestures. Given that no comparison has been conducted yet on meaningless gestures, it would be beneficial to test if the visual-motor EEG oscillations are specific to communicative gestures or generalize to other gesture typologies (e.g. non-communicative gestures), and furthermore to all instances of biological motion, regardless of communicative intent.

The current findings are consistent with the premise that inflection points are salient characteristics involved in gesture perception, recognition and repetition. As such, the oscillation found in EEG mu rhythms are sensitive to occurrences of inflection points during gesture processing. The cognitive processes related with gesture production and perception could potentially be considered as a source of features for gesture recognition.

## 4. Conclusions

The purpose of this paper is to validate the claim that an executed gesture contains a finite set of salient points that produce neural signatures, involving oscillations in EEG signals, associated with cognitive processes. Those signatures may be used to keep in memory intrinsic characteristics of the gesture. This was achieved by determining whether a relationship exists between the time occurrences of salient events in both motion trajectories of performed gestures, and mu oscillatory rhythms for participants observing the gestures.

A statistical analysis was conducted using linear regression using three different conditions. Peaks in the EEG signal at central electrodes and occipital electrodes were used to isolate the salient events within each gesture. Average lags in EEG power oscillations were detected at 343ms and 400ms after inflection points at occipital and central electrodes, respectively. The results suggest that coordinated activity in the visual and motor cortices are highly correlated with key motion components within gesture trajectories, and it is consistent with the proposal that inflection points operate as placeholders in gesture recognition. The potential of this work is that it provides evidence that inflection points are key points in gesture trajectories, which can encapsulate the gestures in human cognition. Therefore, these points can be used to capture large variability in each gesture while keeping the main traits of the gesture class.

## 5. Material and Methods

The overview of the proposed experiment can be found in

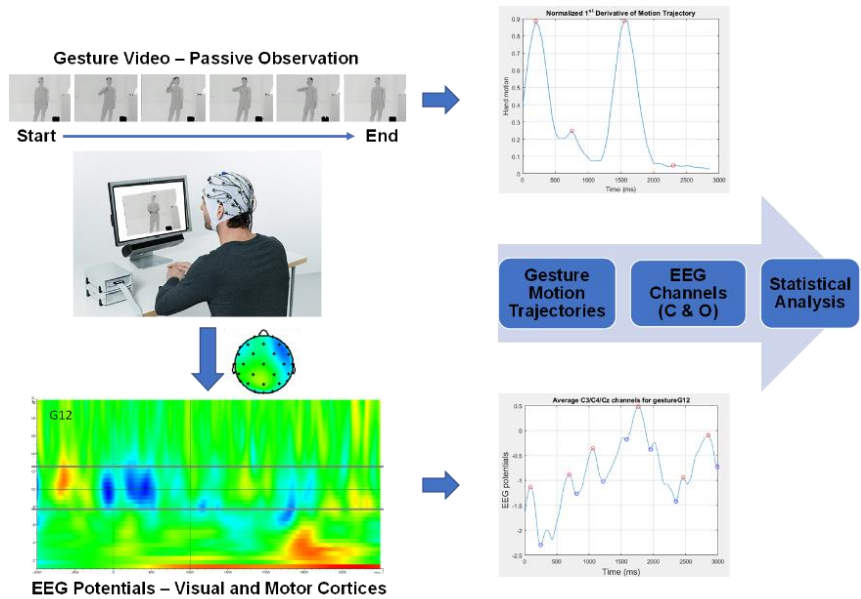

Figure 7. Using recorded videos from a gesture vocabulary, EEG data was acquired from participants watching the gesture movements. The gesture videos were processed to gather motion information, gathering information from salient points in motion. EEG data was processed, and peak occurrences information is extracted. These two data sources were used in statistical analyses, to check whether a relationship exists between salient kinematic points in gesture production and neural signatures found during gesture perception.

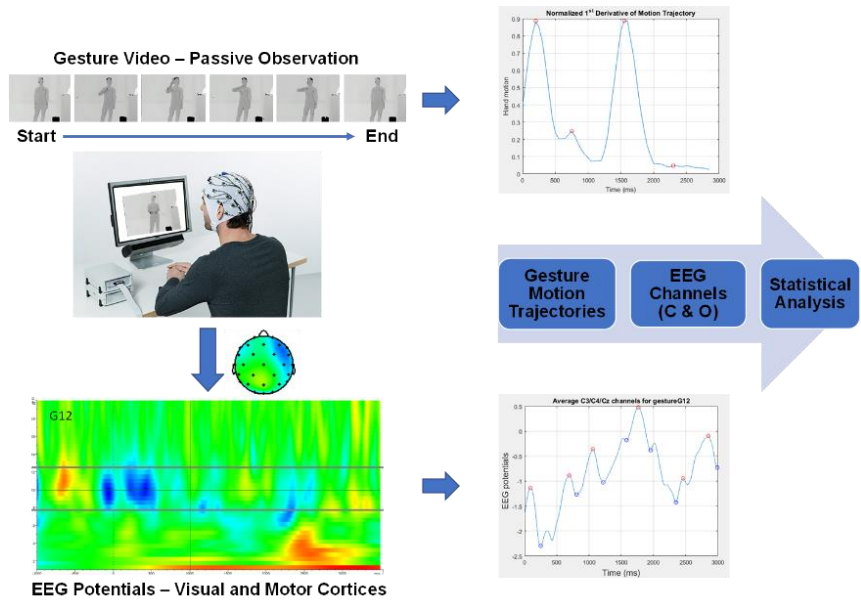

*Figure 7.* Overview of the proposed experiment. Motion trajectories were extracted from each gesture to identify inflection points. Continuous EEG was collected while participants passively viewed the gestures, and time-frequency analyses were used to isolate the time course of mu oscillations in response to each gesture.

### 5.1. Participants

The data presented was collected from seventeen adult volunteers, as part of a larger ongoing project (Age: *M*=21, *SD*=4, Range=18–34; Gender: 11 female, 6 male). All participants gave written consent, with ethical approval given by Purdue University Institutional Review Board. Preliminary analyses of four participants were published previously (Cabrera et al., 2017).

## 5.2. Gesture Task

Participants passively viewed 20 videos taken from the ChaLearn 2013 stimulus set (Escalera et al., 2013). This data set is comprised by 20 gestures from Italian signs, which are communicative in nature ( 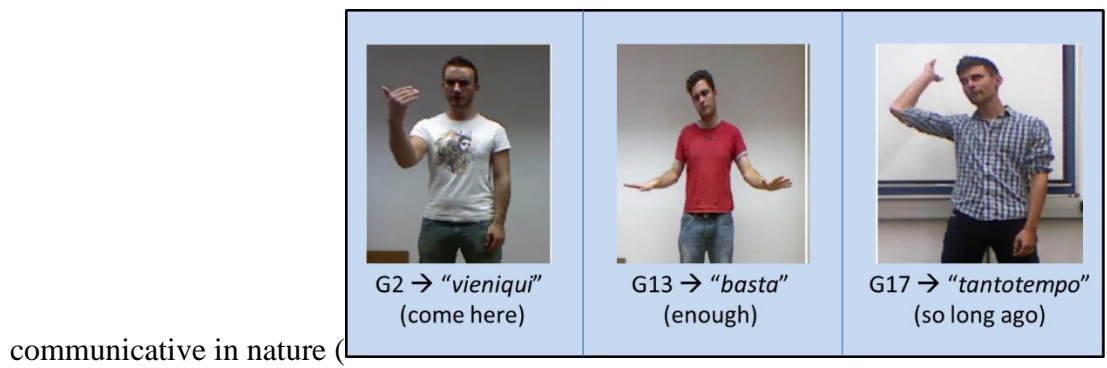

Figure 8). These hand gestures contain anthropological and cultural significance. There are approximately 14000 gestures acquired from 30 different users. For this particular experiment, only one instance of each gesture class was used, and they were separated into individual video files. Each video was approximately 3 seconds in length and showed a single actor performing a gesture. Videos were presented in black and white and only the silhouette of the actor was visible, thereby making the physical motion of the gesture the most salient aspect of the video. Participants viewed each gesture a total of six times, with an inter-stimulus interval of 2 seconds; gesture order was randomized across participants.

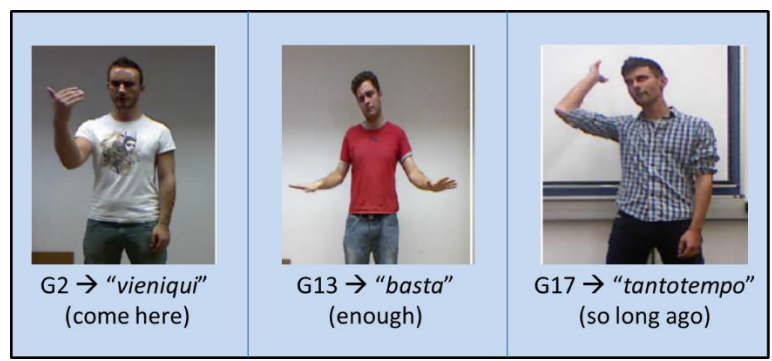

*Figure 8.* Examples of communicative gestures found in the ChaLearn 2013 lexicon

### 5.3. Extracting Motion Information from Gesture Videos

The ChaLearn 2013 data set includes information regarding the position of 25 joints in the human body with respect to the sensor's reference frame, additionally to the color and grayscale videos. With this information, it is possible to identify the location of the user's hands in every frame of the selected gesture videos. The changes in position for both hands along the time axis (using all the frames in the video sequence) are used to describe the trajectory of the hands during the gesture execution.

By taking the derivative of both hands' motion trajectory, it was possible to determine the occurrence of inflection points in motion. The inflection points are the local maxima or minima found in the derivative of the hands position over time. These are associated with abrupt changes in speed and orientation, which define the transition between concatenated movement phases (Kita et al., 1998). These abrupt changes that divide the gesture motion into phases are not dependent on the duration of the gesture (Viviani and Terzuolo, 1982), but they are bounded by a sequential order (Bobick and Wilson, 1997). The use of inflection points (IP) has been found in previous literature and associated with depicting the gesturer's intentions (Buchin et al., 2011; Despinoy et al., 2016; Loram et al., 2006).

Following this reasoning based on previous literature, inflection points are considered a salient characteristic or "placeholders" within each gesture class, under the assumption that the same placeholders would appear regularly across multiple examples of the same gesture class, and independently of variability associated with human nature. A different assumption tested, and concerning this work, is that such characteristics are highly correlated to the neural responses during the process of gesture perception.

The placeholders for each gesture class were obtained in the following way: the position of both hands was extracted for each frame of the video and the derivative of the motion was calculated. Then, the time occurrences for local maxima and minima on the derivative signal were extracted. By doing a transformation between the frame number where the inflection occurred, and the rate of frame per seconds (20fps) of the videos, the occurrence of inflection points was determined for each gesture class.

The detailed information of the ChaLearn 2013 gesture sample used as stimuli for the experiment can be found in Appendix A, including the time stamps of the extracted inflection points in the gestures' motion trajectory. Overall, the extracted motion information can be summarized as follows:

- All gesture classes contained more than one inflection point. There were 3 gesture classes with only 2 IP, namely "G2", "G19", and "G20".
- 17 out of the 20 gesture classes (85%) displayed at least 3 IP, of which seven contained 4 IP and one ("G6") contained 5 IP.
- Across all gesture classes, a total of 66 IP were extracted and correlated with EEG data.

**5.4. EEG Acquisition and Analysis**

Continuous EEG responses were recorded using an ActiCap and the ActiCHamp amplifier system (Brain Products). The signal was digitized at 24-bit resolution and a sampling rate of 500 Hz. Recordings were taken from 32 scalp electrodes based on the 10/20 system, with a ground electrode at Fpz. BrainVision Analyzer (Brain Products) was used for offline analysis. Data were referenced to the average two mastoids (TP9/TP10) and bandpass filtered from 0.1-100 Hz. EEG signals were segmented from -1000 to 5000ms relative to video onset, and ocular correction was performed using a regression-based algorithm. Artifacts were rejected within individual channels using an automated procedure: A 50 µV per ms voltage step with a maximal difference of 100 µV or more within 200 ms, and lowest activity of .5 µV within 100 ms.

Mu oscillations in response to each gesture were quantified using wavelet analysis. As opposed to a Fourier transform, which assumes a time invariant signal, wavelet transforms can be used to quantify changes in power over time. Briefly, a wavelet function at a specified frequency with a specified number of cycles was convolved with the EEG signal in order to quantify fluctuations in power at that frequency (Herrmann et al., 2005). Here, complex Morlet wavelets were applied separately to each gesture observation, using a frequency range of 1-20 Hz, frequency steps of 0.5 Hz, and Morlet parameter of c=5. Segments were baseline corrected relative to the -500 to -100ms window and averaged separately for each gesture type (i.e., averaged across the 6 viewings of each gesture). Analyses focused on the wavelet layer peaking at 10 Hz, corresponding to mu/alpha activity. Each layer provided a time series indicating relative power at 10 Hz at that time point and for that electrode. Peaks within this time series were then calculated in order to examine the time course of mu oscillations. We extracted wavelet layer at C3/Cz/C4 (i.e., directly above the motor cortex), where mu activity is known to be maximal. For comparison, we also considered O3/Oz/O4 (i.e. an occipital cluster). We expected that these occipital electrodes would capture visual cortical activation associated with stimulus processing but not mu activity per se (Streltsova et al., 2010).

**5.5. Statistical Analysis**

For this analysis, the three channels associated with the motor cortex responses (C3/Cz/C4) were averaged over and will be further referred to as Central Cluster; analogously, the channels associated with visual cortical activation (O3/Oz/O4) are further referred to as the Occipital Cluster. Inflection points extracted from one ChaLearn's lexicon, are the local maxima found in the first derivative for the motion trajectory of the gesturer's hands for each gesture class. Time stamps of inflection points from motion data and peak information from EEG data were plotted (time vs. time). Trend lines and coefficients of determination were obtained to check correlation on the plotted data. Additional linear regression models were fitted, and the models' coefficients were determined to check the significance of their predictive nature. The aforementioned metrics were used in three different analyses:

*Standard Analysis:*

For this analysis, the underlying assumption is that salient occurrences in the gesture motion trajectory, hypothesized to be the inflection points, generate a subsequent peak in neural response as 10 Hz oscillations in the EEG data. The regression models were fitted to determine the timing of peaks in EEG data using the timing of inflection point occurrence as the predictors.

*Reverse Analysis*

This analysis considers the opposite assumption: peaks in EEG data precede occurrences of inflection points in gesture motion. Given the nature of the regular analysis, this analysis performs a different measurement with the intention of finding an alternative significance of the proposed scenario. If this analysis shows no meaningful correlation, it would discourage the possibility of assigning the correlations found in the regular analysis as spurious relationships. The regression models were fitted to determine the occurrence of inflection points using EEG peak data as the predictors.

*Random Analysis*

For this analysis, random EEG data was considered under the same assumption as the regular analysis, occurrences in gesture motion precede EEG peaks. In order to achieve this, the same EEG acquired data was flipped in time and motion information of gesture class $i$ was paired with EEG data from gesture class $j$, where $i \neq j$. By shuffling EEG data, random oscillations were achieved while maintaining realistic variability within the data. The purpose of this manipulation of the EEG data, is to determine whether a relationship would arise from the analysis of inflection points used as predictors of EEG oscillations, if the EEG data was random, without generating "synthetic" EEG data.

*Multi-level Hierarchical Regression*

This analysis is proposed as an alternative check of the influence of inflection point occurrences in motion as predictors for EEG peaks, using the data from the regular and reverse analysis. The hierarchical regression allows to modify the predictors to generate a model and check the significance of the correlation and the coefficients within the models. The steps in the multi-level regression were as follows: first, only information from the reverse analysis in motion was used to predict EEG peaks, then a model was fitted including the motion information from both the regular and reverse analysis. The SPSS software was used for the statistical analysis, and residuals and normality assumptions were checked.

# Appendix A

A supplementary document includes information regarding the selected video file with the gesture instances from the ChaLearn data set. It includes information about all inflection points found and their frame number related to time occurrence.